\documentclass[12pt]{article}
\usepackage{graphicx}
\newcommand{\Frac}[2]{\frac{\displaystyle #1}{\displaystyle #2}}
\newcommand{\beq}{\begin{equation}}
\newcommand{\eeq}{\end{equation}}
\newcommand{\beqn}{\begin{eqnarray}}
\newcommand{\eeqn}{\end{eqnarray}}
\newcommand{\beqns}{\begin{eqnarray*}}
\newcommand{\eeqns}{\end{eqnarray*}}

\begin{document}

\begin{titlepage}
\begin{center}

\hfill USTC-ICTS-17-10\\
\hfill September 2017

\vspace{2.5cm}

{\large {\bf  Rare weak decays of $\eta^\prime\to K\pi$}}\\
\vspace*{1.0cm}
 {Dao-Neng Gao$^\dagger$ \vspace*{0.3cm} \\
{\it\small Interdisciplinary Center for Theoretical Study,
University of Science and Technology of China, Hefei, Anhui 230026
China}}

\vspace*{1cm}
\end{center}
\begin{abstract}
\noindent
Rare weak decays of $\eta^\prime\to K\pi$ have been investigated in the framework of the $U(3)$ chiral perturbation theory at the leading order.  Our study shows that the branching ratio ${\cal B}(\eta^\prime\to K\pi)$ is of the order of $10^{-11}$, which is far below the present experimental upper bound given by the BESIII Collaboration. By further analysis of $\eta^\prime\to K^+\pi^-$ and $\eta^\prime\to K^0\pi^0$, the ratio of isospin amplitudes is found that $|A_{1/2}/A_{3/2}|\simeq 35$, which supports that the $\Delta I=1/2$ transition enhancement, namely, the $\Delta I=1/2$ rule, could be functional in $\eta^\prime$ weak decays.

\end{abstract}

\vfill
\noindent
$^{\dagger}$ E-mail:~gaodn@ustc.edu.cn
\end{titlepage}

Nonleptonic weak decays for light hadrons, such as Kaon decays $K\to \pi\pi$, provide a very useful laboratory to investigate low energy dynamics of the standard model (SM). Very recently, a rare but interesting weak decay, $\eta^\prime \to K^\pm \pi^\mp$, was studied experimentally for the first time by the BESIII Collaboration \cite{BES}, and the upper bound for the ratio
\beq\label{upperbound}\frac{{\cal B}(\eta^\prime\to K^\pm \pi^\mp)}{{\cal B}(\eta^\prime\to \gamma \pi^+\pi^-)}<1.3\times 10^{-4}
\eeq
was obtained.  Using the observed branching ratio of $\eta^\prime \to \gamma \pi^+\pi^-$ \cite{PDG2016}, we have
\beq\label{br}{\cal B}(\eta^\prime\to K^\pm \pi^\mp)<3.8\times 10^{-5}.
\eeq

Theoretically, nearly thirty years ago, a very crude estimate for the decay rate of this transition was performed in Ref. \cite{BR88} by assuming the $\Delta I=1/2$ transition enhancement and a simple mass rescaling as
\beq\label{BR1988}
\Gamma(\eta^\prime\to K\pi)\approx (m_{\eta^\prime}/m_K)^3 \Gamma(K_S\to \pi\pi),\eeq
which gives a branching ratio of the order of $10^{-10}$. It has been pointed out in \cite{BR88} that the uncertainty in this estimate is quite large.

The so-called $\Delta I=1/2$ rule was first established in $K\to \pi\pi$ decays, and its origin still remains elusive to this day. Due to Bose symmetry, the two-pion final state from the decay of a $K$ meson can only be in a state of isospin $I=0$ or $I=2$, thus the total decay amplitude can be decomposed into the isospin amplitudes denoted as $A_0$ or $A_2$, respectively. Since a Kaon state has isospin $I=1/2$, $A_0$ is dominated by $\Delta I=1/2$ transitions and $A_2$ gets contributions from $\Delta I=3/2$ transitions. Experimentally, it is shown that \cite{PDG2016, BGB14}
\beq\label{deltaI}
\left|\frac{A_0}{A_2}\right| \approx 22,\eeq
which indicates that the $\Delta I=1/2$ transitions are strongly enhanced.

Similarly, note that the $K\pi$ final state could have isospin $3/2$ and $1/2$, and $\eta^\prime$ is an isoscalar,  the $\Delta S=1$ weak nonleptonic $\eta^\prime\to K\pi$ decays receive contributions from both the $\Delta I=1/2$ and $\Delta I=3/2$ parts of the weak Hamiltonian \cite{BR88}, which can be parameterized as
\beqn\label{isospinamplitde1}
A(\eta^\prime\to K^+\pi^-)=\sqrt{\frac{1}{3}} A_{3/2}-\sqrt{\frac{2}{3}}A_{1/2},\\\label{isospinamplitde2}
A(\eta^\prime\to K^0\pi^0)=\sqrt{\frac{2}{3}} A_{3/2}+\sqrt{\frac{1}{3}}A_{1/2}.
\eeqn
 Here $A_{3/2}$ and $A_{1/2}$  are the corresponding isospin amplitudes, and in general they are complex due to the strong phase. In the limit of CP invariance, $\Gamma(\eta^\prime\to K^+\pi^-)=\Gamma(\eta^\prime\to K^-\pi^+)$ and $\Gamma(\eta^\prime\to K^0\pi^0)=\Gamma(\eta^\prime\to \bar{K}^0\pi^0)$, hence we do not need to consider these CP conjugate processes here. Now one can realize that, besides Kaon decays,  studies of $\eta^\prime\to K \pi$ decays may also provide some possibilities to increase our understanding of the $\Delta I=1/2$ rule. Furthermore, as rare decay modes, $\eta^\prime\to K\pi$ transitions, which should be suppressed in the SM, however, might get enhancement in some novel scenarios. Thus these studies could also be very helpful to explore new physics beyond the SM.

 The main purpose of the present paper is devoted to the analysis of $\eta^\prime\to K\pi$ decays in the SM. Due to the non-perturbative nature of strong interactions at low energy, weak dynamics for light mesons like nonleptonic Kaon decays is generally described in the framework of chiral perturbation theory ($\chi$PT) \cite{Ecker95, Pich95, DI98}. In order to include $\eta^\prime$ mesons, one should extend the standard $SU(3)$ $\chi$PT to the $U(3)$ case through the large-$N_c$ approach, with $N_c$ the number of Quantum Chromodynamics (QCD) colors, and such chiral lagrangian in the strong sector has been presented in \cite{Leu96,HLPT, Pich98, KL00} (actually, the leading order chiral lagrangian for nonleptonic weak interactions including $\pi$, $K$, $\eta$ and $\eta^\prime$ mesons has been studied by the authors of Refs. \cite{GST05, GM12}). In the large $N_c$ limit, the singlet axial current is also conserved and $U(1)_A$ anomaly is absent, thus the QCD lagrangian will have a larger $U(3)\times U(3)_R$ chiral symmetry, so that the $SU(3)$ singlet $\eta_0$ also becomes the Goldstone boson, which can be systematically incorporated into the $U(3)$ chiral lagrangian. The spontaneous symmetry breaking of $U(3)_L\times U(3)_R \to U(3)_V$ gives then rise to a nonet of pseudoscalar Goldstone bosons, which, using the conventional exponential parametrization, can be collected in a unitary $3\times 3$ matrix $U$ in flavor space as follows \cite{Leu96,HLPT, Pich98, KL00}
 \beqn\label{U}
 &&U={\rm exp}(i\sqrt{2}\Phi/F),\\\nonumber\\\label{PHI}
 &&\Phi=\left(\begin{array}{ccc}
 \Frac{\pi^0}{\sqrt{2}}+\Frac{\eta_8}{\sqrt{6}}+\Frac{\eta_0}{\sqrt{3}} & \pi^+ & K^+\\
 \pi^-& -\Frac{\pi^0}{\sqrt{2}}+\Frac{\eta_8}{\sqrt{6}}+\Frac{\eta_0}{\sqrt{3}}& K^0\\
 K^- & \bar{K}^0 & -\Frac{2 \eta_8}{\sqrt{6}}+\Frac{\eta_0}{\sqrt{3}}\\
 \end{array}
 \right),
 \eeqn
and $F=92.4$MeV is the pion decay constant. Thus, as given in Refs. \cite{GST05, GM12}, the ${\cal O}(p^2)$ effective Hamiltonian guiding the $\Delta S=1$ nonleptonic weak interactions can be expressed as
\beq\label{effectiveH}
{\cal H}_W^{\Delta S=1}=G_8 Q_8+ G_8^s Q_8^s+G_8^m Q_8^m+G_{27} Q_{27} +{\rm H.c.},
\eeq
where $G_8$, $G_8^s$, $G_8^m$, and $G_{27}$ are effective couplings, the standard operators
\beqn\label{Q8}
&&Q_8=(L_\mu L^\mu)_{23},\\
&&Q_8^s=(L_\mu)_{23}\langle L^\mu\rangle,\\
&&Q_{27}=(L_\mu)_{23}(L^\mu)_{11}+\frac{2}{3}(L_\mu)_{13}(L^\mu)_{21}-\frac{1}{3}(L_\mu)_{23}\langle L^\mu\rangle,
\eeqn
are built up from the left-handed currents $L_{\mu}=iF^2 \partial_\mu U U^\dagger$, and $\langle \rangle$ denotes a trace over flavors. It is known that the weak mass term $Q_8^m$ does not contribute to the leading order on-shell amplitudes, while its higher order effects can be absorbed into a redefinition of the weak counterterms \cite{Crew86,KMW90}, we will therefore neglect it hereafter.

As mentioned above, as the ninth Goldstone boson, the singlet $\eta_0$ can be included into the $U(3)$ $\chi$PT, through the exponential realization of $U$ in eq. (\ref{U}). Due to the $SU(3)$ symmetry breaking, the octet $\eta_8$ and the singlet $\eta_0$ will mix each other and generate the two physical states, the $\eta$ and $\eta^\prime$. The $\eta-\eta^\prime$ mixing is an interesting topic in hadron physics. At the leading order in $\chi$PT, since there is only one mixing term from the mass sector, it is suitable to introduce the single mixing angle $\theta_P$ \cite{GST05, GM12}, which relates the $SU(3)$ eigenstates $(\eta_8, \eta_0)$ and the mass eigenstates $(\eta, \eta^\prime)$ as
\beqn\label{mixingangle}
\left(\begin{array}{c}\eta \\\eta^\prime\\\end{array}\right)=\left(\begin{array}{cc}\cos\theta_P & -\sin\theta_P\\\sin\theta_P & \cos\theta_P\\
\end{array}\right)\left(\begin{array}{c}\eta_8\\\eta_0\\\end{array}\right)
\eeqn
This is the usual and simple approach for $\eta-\eta^\prime$ mixing. However, when one would like to perform the higher order calculation, these contributions will not only generate the mixing of the mass term but also the mixing of the kinetic term, thus the two-mixing-angle description scheme is required in general \cite{SSW93,Leu97, AG98, FK98,KL98, GMR01}. For the purpose of the present work, we will adopt the simple approach (\ref{mixingangle}). The mixing angle $\theta_P$ is determined around $-20^\circ$ phenomenologically \cite{Leu96, HLPT}. In the following numerical calculation, we will allow $\theta_P$ to vary inside the range $[-15^\circ, -25^\circ]$, as discussed in Ref. \cite{GST05}.

Now it is straightforward to derive the leading order decay amplitudes of $\eta^\prime\to K\pi$ from eq. (\ref{effectiveH}), which reads
\beqn\label{amp1}A(\eta^\prime\to K^+\pi^-)=\sqrt{2}F\left(\frac{G_8}{\sqrt{6}}\left[(3 m_{\eta^\prime}^2-2 m_K^2-m_\pi^2)\sin\theta_P+ 2\sqrt{2}(m_\pi^2-m_K^2)\cos\theta_P\right] \nonumber \right.\\
\left. +\sqrt{3} G_8^s(m_\pi^2-m_K^2)\cos\theta_P+\frac{G_{27}}{\sqrt{6}}(2 m_{\eta^\prime}^2-3 m_K^2+m_\pi^2)\sin\theta_P\right), \\
\label{amp2}
A(\eta^\prime\to K^0\pi^0)=-\sqrt{2}F\left(\frac{G_8}{2\sqrt{3}}\left[(3 m_{\eta^\prime}^2-2 m_K^2-m_\pi^2)\sin\theta_P+ 2\sqrt{2}(m_\pi^2-m_K^2)\cos\theta_P\right] \nonumber \right.\\
\left. +\sqrt{\frac{3}{2}} G_8^s(m_\pi^2-m_K^2)\cos\theta_P-\frac{G_{27}}{2\sqrt{3}}(3 m_{\eta^\prime}^2-2 m_K^2-m_\pi^2)\sin\theta_P\right).
\eeqn
Thus it is easy to express the decay rate of these processes as
\beq\label{rate}
\Gamma(\eta^\prime\to K\pi)=\frac{1}{16\pi m_{\eta^\prime}}\lambda^{1/2}(1, r_K,r_\pi) |A(\eta^\prime\to K\pi)|^2,
\eeq
 where $r_K=m_K^2/m_{\eta^\prime}^2$, $r_\pi=m_\pi^2/m_{\eta^\prime}^2$, and $\lambda(a,b,c)=a^2+b^2+c^2- 2(ab+ac+bc)$. Meanwhile, using eqs. (\ref{isospinamplitde1}) and (\ref{isospinamplitde2}) together with decay amplitudes of eqs. (\ref{amp1}) and (\ref{amp2}), one can obtain the ${\cal O}(p^2)$ $\eta^\prime\to K \pi$ isospin amplitudes as follows
\beqn\label{A3/2}
A_{3/2}&=&\frac{5F}{3}G_{27} (m^2_{\eta^\prime}-m_K^2)\sin\theta_P,\\
\label{A1/2}
A_{1/2}&=&-F\left(\frac{G_8}{\sqrt{2}}\left[(3 m_{\eta^\prime}^2-2 m_K^2-m_\pi^2)\sin\theta_P+ 2\sqrt{2}(m_\pi^2-m_K^2)\cos\theta_P\right] \right.\nonumber\\
&&\left. +3 G_8^s(m_\pi^2-m_K^2)\cos\theta_P-\frac{G_{27}}{3\sqrt{2}}(4 m_K^2-m_{\eta^\prime}^2-3 m_\pi^2)\sin\theta_P\right).
\eeqn
 It is easy to see that the $\Delta I=3/2$ transition is only induced by the 27-plet operator $Q_{27}$, which is expected and the same as the case of $K\to\pi\pi$ decays, while the $\Delta I=1/2$ piece receives contributions from both the octet and the 27-plet operators including $Q_8$, $Q_8^s$, and $Q_{27}$.

Our next task is to evaluate the magnitude of the decay rates and check whether the $\Delta I=1/2$ rule is functional or not in these decays. However, the effective couplings $G_8$, $G_8^s$, and $G_{27}$ are unknown constants in the amplitudes, and theoretically, we have no model-independent way to fix them reliably. One should appeal to the phenomenological determination through the experimental input. Generally, $G_8$ and $G_{27}$ can be extracted from the data of $K\to \pi\pi$ decays (in the isospin limit, the $Q_8^s$ piece is absent in the amplitudes), which gives \cite{Ecker95, Pich95, DI98}
\beq\label{G8num}
G_8=9.1\times 10^{-6}~{\rm GeV}^{-2},
\eeq
and
\beq\label{G27num}
\frac{G_{27}}{G_8}\simeq\frac{1}{18}.
\eeq
The weak coupling $G_8^s$, related to dynamics of the singlet $\eta_0$, is peculiar to the $U(3)$ framework. Interestingly, phenomenological constraints on it, from a number of radiative $K$ decays involving pseudoscalar pole diagrams, has been carefully investigated by the authors of Ref. \cite{GST05}, and the ratio
\beq\label{G8snum}\frac{G_8^s}{G_8}\simeq -\frac{1}{3}\eeq
has been obtained. In particular, from $K_L\to\gamma\gamma$, for the mixing angle $\theta_P$ inside the range $[-15^\circ, -25^\circ ]$, the extracted value of $G_8^s/G_8$ is from $-0.35$ to $-0.25$ \cite{GST05}.

\begin{table}[t]\begin{center}\begin{tabular}{ c|c| c| c| c} \hline\hline
 $\theta_P$ & $G_8^s/G_8$ & ${\cal B}(\eta^\prime\to K^+\pi^-)$ & ${\cal B}(\eta^\prime\to K^0\pi^0)$ &$|{A_{1/2}}/{A_{3/2}}|$ \\\hline
 $-15^\circ$ & $-0.35$&$1.4\times 10^{-11}$ &$6.2\times 10^{-12}$ &$38.4$ \\
$-17.5^\circ$& $-0.32$ & $1.8\times 10^{-11}$ &$7.9\times 10^{-12}$ &$37.5$ \\
$-20^\circ$& $-0.29$& $2.2\times 10^{-11}$& $9.9\times 10^{-12}$& $36.8$ \\
$-22.5^\circ$&$-0.27$ & $2.7\times 10^{-11}$&$1.2\times 10^{-11}$& $35.9$\\
$-25^\circ$&$-0.25$&$3.1\times 10^{-11}$&$1.4\times 10^{-11}$&$35.3$\\\hline
\hline
\end{tabular}\caption{Branching ratios of $\eta^\prime\to K\pi$ decays for different values of the ratio $G_8^s/G_8$. The range of the mixing angle $\theta_P$ and the ratio of $G_8^s/G_8$ are taken from Ref. \cite{GST05}, extracted from $K_L\to\gamma\gamma$ decay.} \end{center}\end{table}

 Using the above inputs, it is easy to illustrate our numerical results for the present study, which has been displayed in Table 1. It is seen that our prediction for the branching ratio of $\eta^\prime\to K\pi$ decays is around $10^{-11}$, which is far below the present experimental upper bound by the BESIII Collaboration in eq. (\ref{br}). The ratio of $|A_{1/2}/A_{3/2}|$ for $\eta^\prime \to K\pi$ is found to be from $35.5$ to $38.4$,  comparing with the $K\to\pi\pi$ case, $A_0/A_2 \simeq 22$ given in eq. (\ref{deltaI}). This means that the $\Delta I=1/2$ enhancement also works in $\eta^\prime$ weak decays.

 On the other hand, one can take the relations (\ref{G27num}) and (\ref{G8snum}) together with eqs.(\ref{A1/2}) and (\ref{A3/2}) to estimate this ratio, thus the effective couplings will cancel each other. This leads to
 \beq\label{deltaIetap}
 \left|\frac{A_{1/2}}{A_{3/2}}\right| \simeq 25.5-3.6 \cot\theta_P,\eeq
where the values of the mass of mesons have been used already. Further, for $\theta_P$ around $-20^\circ$, the ratio
\beq
\left|\frac{A_{1/2}}{A_{3/2}}\right|\simeq 35.5
\eeq
is achieved.

To summarize, motivated by the recent experimental study by the BESIII Collaboration, we analyze rare weak decays of $\eta^\prime\to K\pi$ in the framework of the chiral lagrangian for the first time. In the limit of CP symmetry, both of decay amplitudes of $\eta^\prime\to K^+\pi^-$ and $\eta^\prime\to K^0\pi^0$ are examined. It is found that ${\cal B}(\eta^\prime\to K\pi)$ is a few $\times$ $10^{-11}$, far below the current experimental upper limit. The ratio of isospin amplitudes $|A_{1/2}/A_{3/2}|$ has been calculated, which supports that the $\Delta I=1/2$ rule is functional in the weak decays of $\eta^\prime$ meson. The small branching ratio of the decay means a big experimental challenge. Meanwhile, it might also offer interesting room for a new physics probe in the future study.

\vspace{0.5cm}
\section*{Acknowledgments}
This work was supported in part by the NSF of China under Grants No. 11575175 and No. 11235010, and by the CAS Center for Excellence in Particle Physics (CCEPP).

\end{document}